# Magnetic field-enhanced two-electron oxygen reduction reaction using CeMnCo nanoparticles supported on different carbonaceous matrices

Caio Machado Fernandes[1], João Paulo C. Moura[1], Aline B. Trench[1], Odivaldo C. Alves[2], Yutao Xing[3], Marcos R. V. Lanza[4], Júlio César M. Silva[2], Mauro C. Santos[1*]

[1]*Laboratório de Eletroquímica e Materiais Nanoestruturados, Centro de Ciências Naturais e Humanas, Universidade Federal do ABC – Santo André, SP – Brasil*

[2]*Laboratório de Materiais da UFF, Instituto de Química, Universidade Federal Fluminense – Niterói, RJ – Brasil*

[3]*Laboratório de Microscopia Eletrônica de Alta Resolução, Centro de Caracterização Avançada para a Indústria de Petróleo, Universidade Federal Fluminense – Niterói, RJ – Brasil*

[4] *Instituto de Química de São Carlos (IQSC), Universidade de São Paulo (USP), Avenida Trabalhador São-carlense 400, 13566-590, São Carlos, SP, Brazil.*

Corresponding Author:

*E-mail: mauro.santos@ufabc.edu.br (M. C. Santos).

**Abstract**

The current study illustrates the successful synthesis of $Ce_{1.0}Mn_{0.9}Co_{0.1}$ nanoparticles, characterized through XRD, EPR, magnetization curves, and TEM/HRTEM/EDX analyses. These nanoparticles were then loaded into the carbon Vulcan XC72 and the carbon Printex L6 matrices in varying amounts (1, 3, 5, and 10% w/w) via wet impregnation method to fabricate electrocatalysts for the 2-electron ORR. Before experimentation, the material was characterized via XPS and contact angle measurements. The electrochemical results produced significant findings, indicating that the electrocatalysts with the nanostructures modifying both carbon blacks notably augmented currents in rotating ring-disk electrode measurements, signifying enhanced selectivity for $H_2O_2$ production. Moreover, our research underscored the significant impact of Magnetic Field-Enhanced Electrochemistry, employing a constant magnetic field strength of 2000 Oe, on 2-electron ORR experiments. Particularly noteworthy were the observed results surpassing the ones without the magnetic field, demonstrating heightened currents and improved selectivity for $H_2O_2$ production (more than 90 %) facilitated by CeMnCo nanoparticles. These significant findings in electrocatalytic efficiency have practical implications, suggesting the potential for developing more efficient and selective catalysts for the 2-electron ORR.

# 1. Introduction

Among the reactions for a more sustainable and eco-friendlier world, the 2-electron Oxygen Reduction Reaction (ORR) is critical, potentially revolutionizing hydrogen peroxide ($H_2O_2$) production methods. Conventional techniques for synthesizing $H_2O_2$ traditionally depended on methods that substantially adversely affect the environment. Adding to this issue is the considerable problem regarding the storage and transportation of $H_2O_2$, which is plagued with risks. $H_2O_2$ is widely used in applications ranging from disinfection and water treatment to chemical synthesis and environmental remediation, making its safe and efficient production crucial. Consequently, the urgent demand for cleaner, energy-efficient solutions has spurred an investigation of different strategies for generating $H_2O_2$ using electrochemistry via ORR. This approach addresses the energy demands of traditional hydrogen peroxide generation and mitigates the issues related to safety, enabling in-situ electrogeneration [1-6].

Carbonaceous matrices have emerged as pivotal catalysts, with carbon black standing out prominently in this regard. Carbon black's unique structure and composition, characterized by its high surface area, excellent electrical conductivity, and remarkable stability, make it an ideal candidate for catalyzing the 2-electron ORR. Its sp2-hybridized carbon network provides ample active sites for $O_2$ adsorption and subsequent reduction. Moreover, carbon black's inherent properties ensure robust performance and longevity, addressing the critical issues of selectivity and durability in ORR catalysis. Although these qualities underscore the significance of carbon black as an electrocatalyst, researchers understand the necessity of modifying this matrix to enhance both efficiency and selectivity toward $H_2O_2$ electrogeneration [7-10].

In pursuing enhancing electrocatalytic activities for the 2-electron ORR, the synergy between metal oxide nanostructures and carbon black has garnered significant attention over the years. Due to their high surface area, unique electronic properties, and tunable surface functionalities, metal oxide nanostructures can effectively modify carbon black into a highly efficient electrocatalyst for $H_2O_2$ electrogeneration. These metal oxide nanostructures act as promoters, significantly improving the carbon black surface for $O_2$ adsorption/reduction. The intimate integration of metal oxides with carbon black optimizes the electrocatalytic performance. It ensures excellent stability and selectivity during the ORR process, offering a promising pathway for developing sustainable and efficient electrocatalysts [11-14].

Sustained efforts to enhance the electrogeneration of $H_2O_2$ persist by implementing structural and compositional modifications in electrocatalysts. Researchers have diligently explored various avenues, investigating intricate catalyst structure and composition changes to boost ORR efficiency. Despite these extensive investigations, the constraints imposed by conventional catalyst development methods are still a hindrance. The limitations of existing approaches highlight the need for innovative strategies and breakthroughs in materials science and electrochemistry to overcome these barriers and pave the way for more efficient and sustainable ORR processes. Finding novel ways to enhance catalyst performance and understanding the fundamental mechanisms in electrocatalytic reactions are essential for driving advancements in the field and achieving more effective energy technologies [15, 16].

In this context, the recent emergence of magnetic field-enhanced electrocatalysis represents a significant breakthrough in oxygen reduction reaction research. This innovative approach harnesses the power of magnetic fields to enhance electrochemical reactions, offering a promising avenue for improving ORR efficiency. By integrating magnetic field effects, researchers have opened new doors to understanding and optimizing ORR processes, leading to more sustainable and efficient energy conversion technologies and pollutant degradation/mineralization. The synergy between magnetic fields and electrocatalysis holds immense potential for revolutionizing the way electrochemistry is approached, with excellent results already being published, paving the way for cleaner and more advanced technology. As scientists delve deeper into this frontier, the prospects for achieving groundbreaking advancements in ORR continue to grow, offering a new approach for a greener and more energy-efficient world [17-24].

Using the Rotating Ring-Disk Electrode technique, we assessed the ORR activity for cerium-manganese-cobalt oxide nanoparticles supported on carbon Vulcan XC-72 and Printex L6. This evaluation focused on their effectiveness as electrocatalysts for the 2 electrons ORR, both with and without a continuous magnetic field. This innovative study presents pivotal advancement in conventional electrocatalysis, showing that introducing a magnetic field can yield substantial advantages, opening new avenues for research and application in electrochemistry.

2. **Experimental Procedure**

## 2.1. CeMnCo oxide nanoparticles synthesis

The intricate oxide nanoparticles comprising Ce, Mn, and Co (abbreviated as $Ce_{1.0}Mn_{0.9}Co_{0.1}$) were synthesized through a high-temperature hydrothermal co-precipitation method. To achieve this stoichiometric composition, a quantity of 0.1 mol $CeCl_3 \cdot XH_2O$, 0.09 mol $MnCl_2 \cdot XH_2O$, and 0.01 mol $CoCl_2 \cdot XH_2O$ was dissolved in distilled water. This solution was gradually added to a 0.25 mol $L^{-1}$ $Na_2CO_3$ solution to pH 9. The product was treated at 600 K for 3 hours, forming $Ce_{1.0}Mn_{0.9}Co_{0.1}$ complex oxide nanoparticles. It was thoroughly washed with distilled $H_2O$ and absolute EtOH and dried at 353 K.

## 2.2. Electrocatalysts preparation

To produce the $Ce_{1.0}Mn_{0.9}Co_{0.1}$ electrocatalysts, two support matrices such as carbon Vulcan XC-72 and Printex L6, were utilized, with metal oxide loadings ranging from 1% to 10% (w/w). The preparation process employed the wet impregnation method. Initially, $Ce_{1.0}Mn_{0.9}Co_{0.1}$ was mixed with carbon and suspended in milli-Q water. This mixture underwent continuous magnetic stirring for 4 hours. Subsequently, they were dried at 373 K. They were named CeMnCo/VC and CeMnCo/PT, differing in the carbon matrix used (VC for Vulcan XC72 and PT for Printex L6).

## 2.3. Physicochemical characterization

X-ray diffraction (XRD) was performed utilizing a Panalytical model X'Pert Pro-PW3042/10 diffractometer, which was equipped with Cu Kα radiation ($\lambda = 0.1540$ nm) at 40 kV and 40 mA and a solid-state X-Celerator detector. The Transmission electron microscopy (TEM) images were obtained using a JEOL JEM 2100F electron microscope. To discuss wettability, the contact angles of the electrocatalyst were evaluated using a goniometer.

X-ray photoelectron spectroscopy (XPS) analysis used a Scienta Omicron ESCA+ spectrometer with monochromatic Al Kα (1486.7 eV) radiation. Shirley's method enhanced accuracy to eliminate the inelastic background in C 1s high-resolution core-level spectra. Spectral fitting was conducted using the CasaXPS software, employing unconstrained multiple Voigt profile fitting techniques.

Magnetic properties of samples were obtained using Vibrating Samples Measurements (VSM) mode in a Physical Property Measurements System (PPMS) DynaCool, from Quantum Design, and Electron Paramagnetic Resonance (EPR) was performed in an EMX Plus (Bruker) with band cavity-X of 9 GHz at room temperature and ten mW microwave power. The paramagnetic signal in EPR results was fitted using the Easyspin Matlab routine [25]

### 2.4. Oxygen Reduction Reaction (ORR)

Electrochemical experiments employed a potentiostat/galvanostat (Autolab PGSTAT 302N) connected to a rotating ring disc electrode system (RRDE) from Pine Instruments. The working electrode configuration included a glass carbon (GC) disc (0.2475 cm²) coupled with a platinum ring (0.1866 cm²) at a collection factor of N = 0.26. A Pt counter electrode measuring two cm² and a Hg|HgO reference electrode were utilized. The supporting electrolyte used was 1 mol L$^{-1}$ NaOH. Electrocatalysts were deposited onto the GC disc through drop casting. Dispersions of the electrocatalyst in water (2 mg mL$^{-1}$) were homogenized via ultrasonication, and 20 μL of the resulting mixture was applied to the disc electrode surface. After drying, a 20 μL aliquot of a 1:100 Nafion solution (v/v, Nafion: deionized water) was added to the electrode film and dried. Before all electrochemical analyses, the electrolyte was oxygen-saturated for 30 minutes, maintaining consistent flow during duplicate measurements conducted at room temperature with a scan rate of 5 mV s$^{-1}$. These procedures were carried out with and without a constant magnetic field of 2000 Oe at a distance of 1.0 cm from the electrode.

## 3. Results and discussion
### 3.1. Physical characterization

The XRD diffractogram of the obtained $Ce_{1.0}Mn_{0.9}Co_{0.1}$ is shown in Fig. 1. The peaks are related to cubic fluorite $CeO_2$. The slight deviation in the values is due to the interaction with Mn and Co. As no discernible peaks are associated with Mn and Co oxides, it can be inferred that they are either dispersed within the ceria material or exist in an amorphous state [26]. No strong background scattering of amorphous materials is observed in the XRD pattern, indicating that the latter case is unlikely to have occurred in the samples.

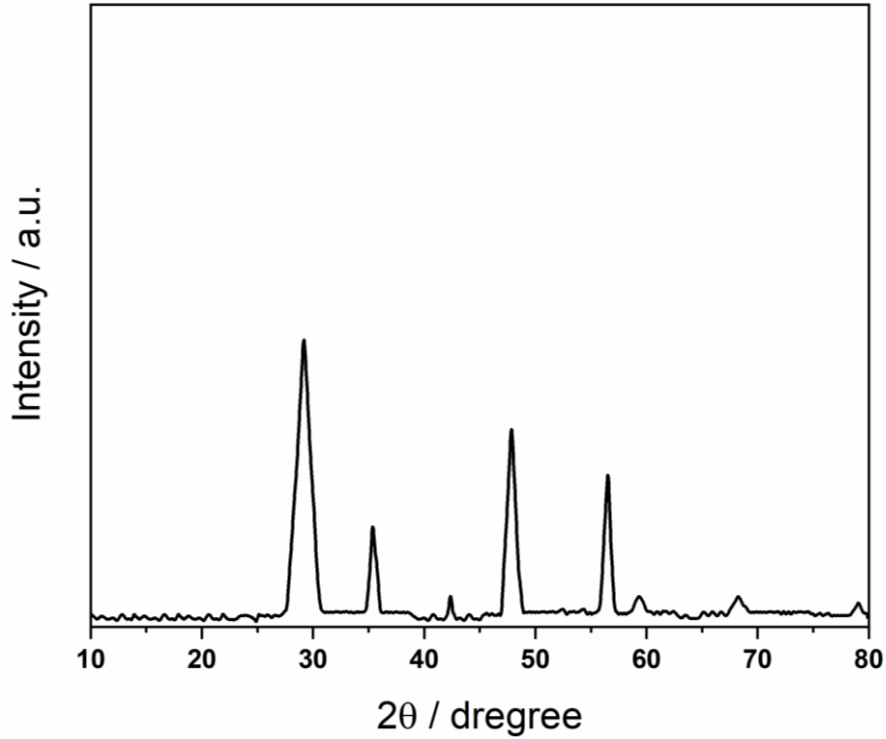

**Figure 1.** XRD pattern of the obtained CeMnCo nanoparticles.

The magnetization (M) versus magnetic field (H) curve at 300 K is depicted in Fig. 2. The result shows a substantial paramagnetic contribution, as shown in Fig. 2a, which was derived through a linear fit of experimental values within the high magnetic field range. Subsequently, the MxH loop, obtained by subtracting the paramagnetic contribution, is displayed in Fig. 2b. Notably, this loop displays a characteristic wasp-waisted behavior near zero applied magnetic field, as illustrated in the inset of Fig. 2b. Such behavior is indicative of a mixture of magnetic phases with differing coercivities, suggesting a combination of ferromagnetism and superparamagnetism [27].

Although Ce atoms do not exhibit ordered magnetic moments, $CeO_2$ displays weak ferromagnetic behavior at room temperature, which is attributed to the presence of deep oxygen vacancies formed on the solid surface [28, 29]. This ferromagnetic property is enhanced in Mn or Co-doped $CeO_2$, where an increased concentration of vacancies near the doping atoms is observed [30]. Given that Co atoms in nanometric dimensions exhibit superparamagnetic behavior [31], it can be inferred that the observed paramagnetism, ferromagnetism, and superparamagnetism are respectively attributed to Mn, Ce, and Co elements.

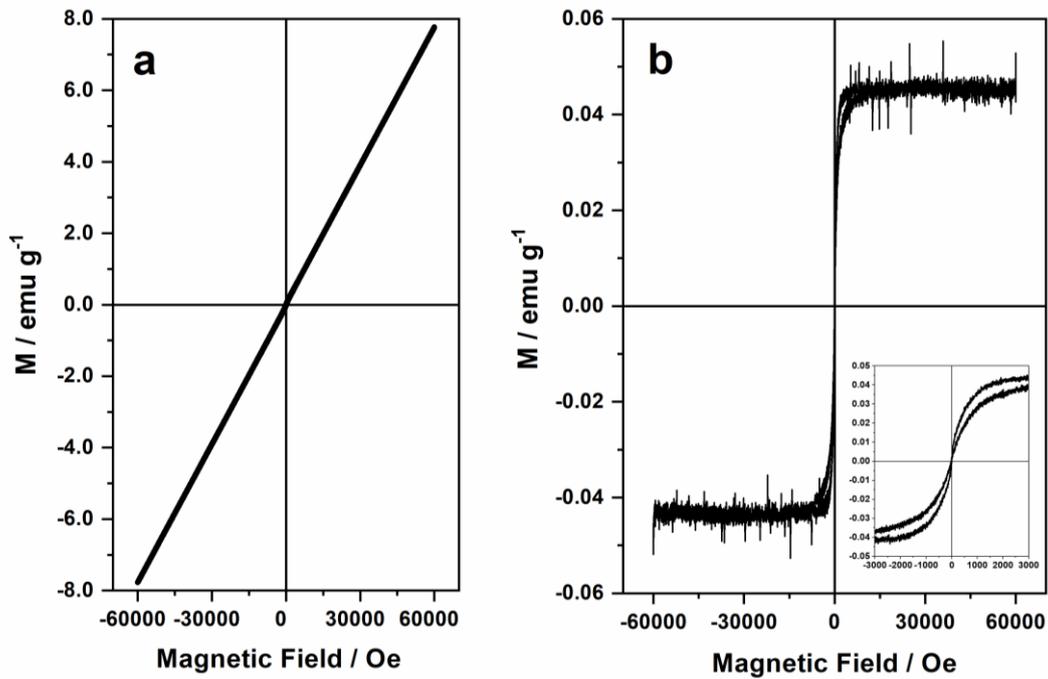

**Figure 2**. Magnetization as a function of the external magnetic field for CeMnCo nanoparticles was measured at 300 K (a), and magnetization for CeMnCo was ordered (b).

Fig. 3 presents the EPR spectrum, a vital tool in our analysis. It reveals two distinct components. The first, a broad line with a g-factor of 1.97 and a linewidth of 2200 Oe, is likely from ferromagnetic or superparamagnetic particles like $CeO_2$ and Co nanoparticles. The second component, visible in the high-field region after subtracting the baseline for the ferromagnetic signal, is displayed in the inset. With six equally spaced lines fitted using the Easyspin routine, this component has a g-factor of 2.00 and a hyperfine splitting constant of A = 90 Oe. This signal arises from the paramagnetic $Mn^{2+}$ ion, characterized by a nuclear spin of I = 5/2 [32]. These findings are crucial in understanding the composition of CeMnCo nanoparticles.

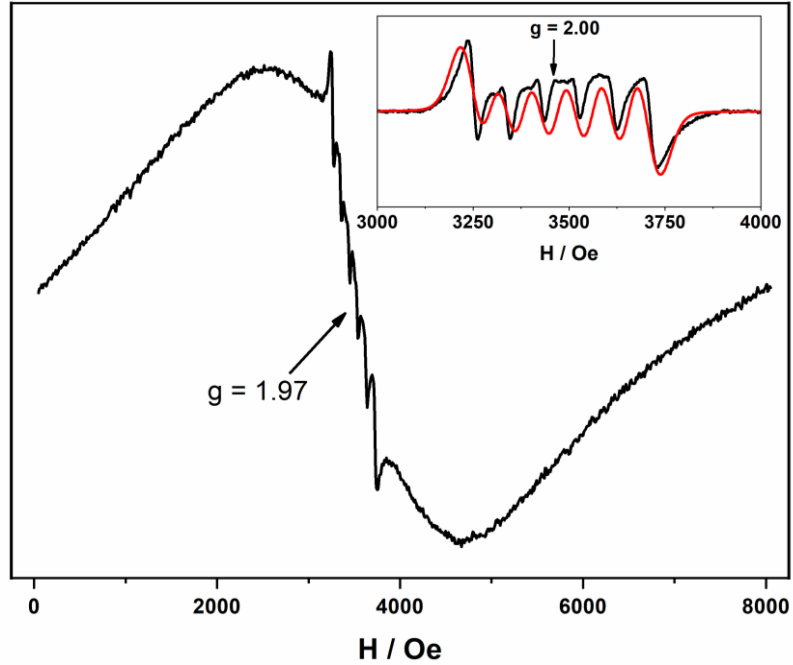

**Figure 3**. EPR spectrum of CeMnCo at room temperature. Insert High field spectra. The red line is the Easypin fit.

TEM micrographs (Fig. 4) of CeMnCo nanoparticles synthesized without controlled morphology reveal a variety of morphologies and size distributions across different magnifications. These nanoparticles exhibit a heterogeneous structure due to the lack of precise synthesis conditions, resulting in diverse shapes and sizes. Despite the morphological variability, the TEM images consistently capture the expected features of CeMnCo nanoparticles, allowing for a comprehensive understanding of their structural characteristics. The role of TEM images in providing this comprehensive understanding cannot be overstated.

The different magnification levels in the TEM images, ranging from 1000 nm to 10 nm, provide valuable insights into the obtained nanostructure, showcasing the overall morphology and distribution and the finer details at higher resolutions, such as spatial arrangements and crystal structure. Also, the energy dispersive X-ray (EDX) mapping analysis (Fig. S1) confirms the presence of Ce, Mn, and Co in the obtained nanostructures, validating the successful synthesis.

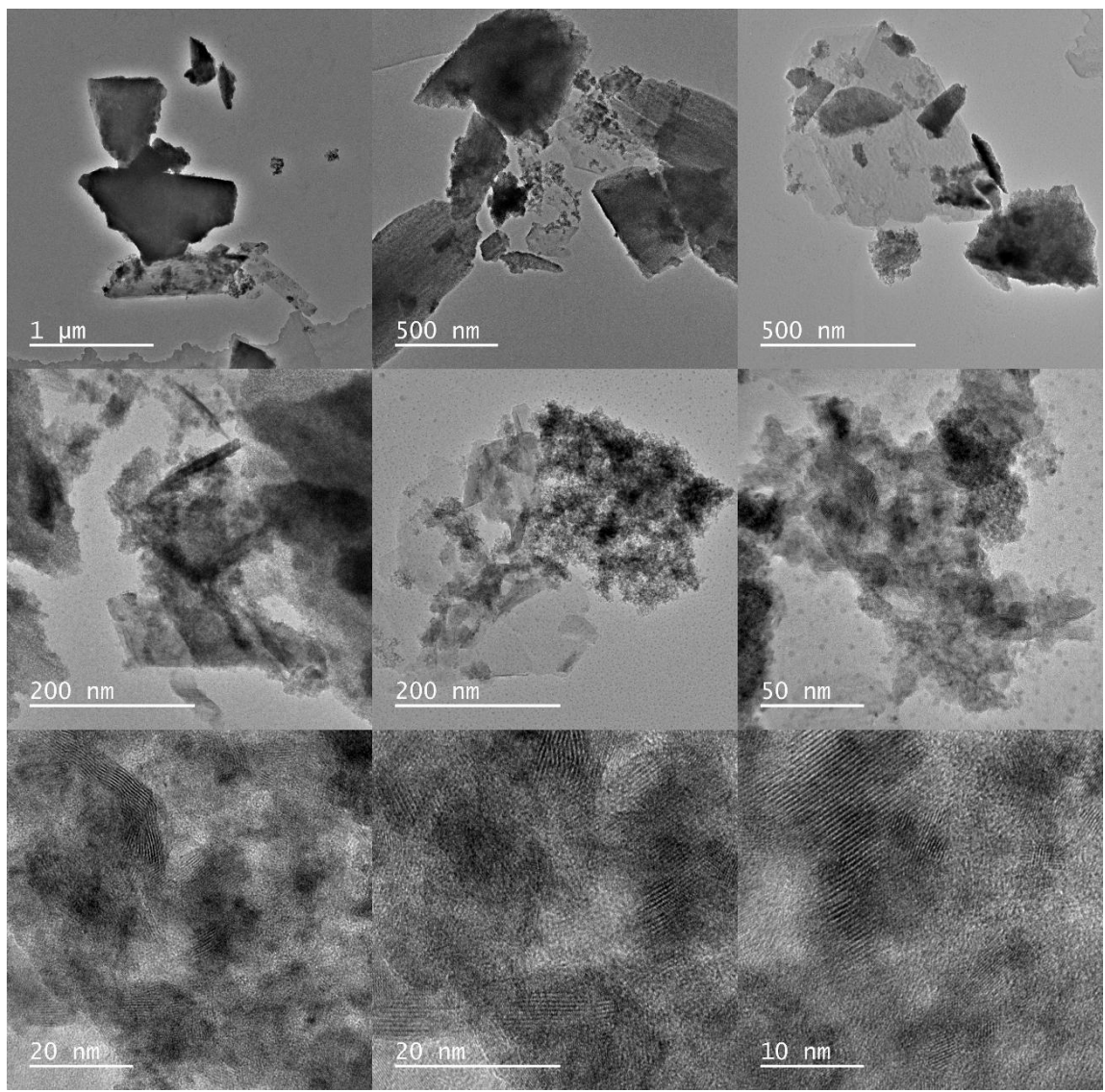

**Figure 4.** Low magnification TEM and HRTEM (lower) images of CeMnCo nanoparticles.

Fig. S2 shows the high-resolution TEM (HRTEM) image and its corresponding Fast Fourier Transform (FFT) of the selected area in red lines. FFT analysis indicated the presence of $CeO_2$, based on the (111) plane with an interplanar distance of 0.309 nm [33, 34], and the presence of CoO, based on the (111) plane with an interplanar distance of 0.248 nm [35], confirming the crystalline structure of the synthesized material.

Fig. 5 displays the XPS C 1s core-level spectra of the Vulcan XC72, Printex L6, 3% CeMnCo/VC, and 3% CeMnCo/PT samples. The spectra were deconvolved into four components, each related to a phase present in carbon. The main component, located at approximately 284 eV, corresponds to the C-C bond. Three other components appear at approximately 285, 286, and 289 eV and are related to the C–O, C=O, and COOH bonds, respectively [36]. Identifying these phases is crucial in validating the successful synthesis of CeMnCo nanoparticles.

In Fig. 5a and 5c, the concentrations of each component are shown for the Vulcan XC72 and Printex L6 samples, respectively. It can be noted that the Vulcan XC72 sample presented 52.20 at.% of oxygenated species, while the Printex L6 sample presented 40.98 at.%. Fig. 5b and 5d also show the concentrations of each component for the CeMnCo/VC and 3% CeMnCo/PT samples, which offered 56.23 at.% and 43.75 at.% of oxygenated species, respectively.

These results indicate that modifying the Vulcan XC72 and Printex L6 carbon materials with CeMnCo nanoparticles promotes an increase in oxygenated species, making the surface more hydrophilic, which can favor the electrogeneration of $H_2O_2$ [37].

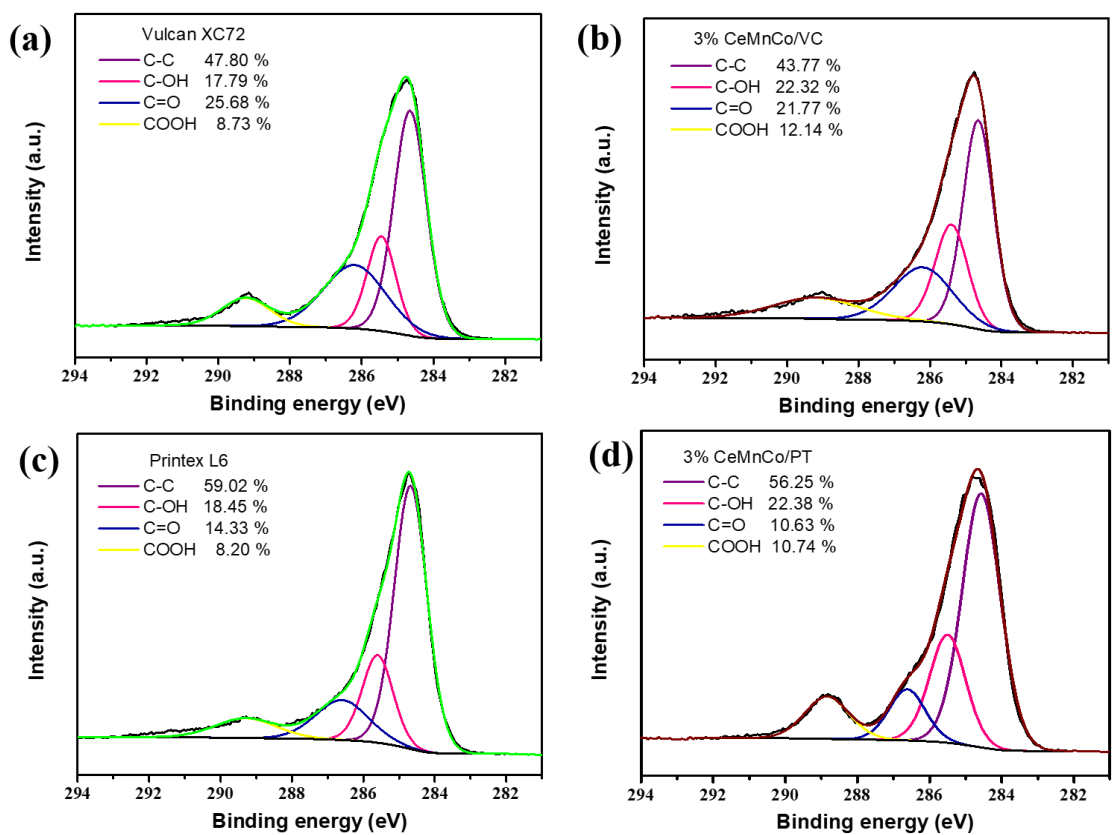

**Figure 5.** Deconvoluted C 1s XPS spectra of (a) Vulcan XC72, (b) Printex L6, (c) 3% CeMnCo/VC, and (d) 3% CeMnCo/PT.

Fig. 6 shows the contact angle measurements. For Vulcan XC72, the obtained value was 46.7 °, and for Printex L6 was 37.4 °. Those values significantly diminished as the CeMnCo nanoparticles were added on both carbon black matrices. They were 28.0 ° for CeMnCo/VC and 29.6 ° for CeMnCo/PT. These findings suggest that the nanostructures made the surfaces more hydrophilic, and the effect can be linked to the rise in oxygenated species, as shown in XPS results, turning both materials into better electrocatalysts for the 2-electron ORR. The hydrophilic nature resulting from abundant oxygen-containing functional groups and the surface texture of the carbon matrix facilitates effective wetting of the catalyst by the aqueous electrolyte Enhancing wettability through this modification improved contact of the active adsorption sites with water and thus facilitated the mass transport of oxygen [38-40].

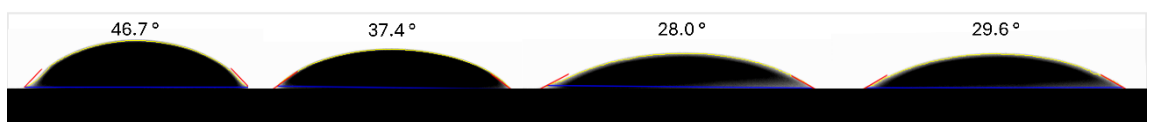

**Figure 6.** The contact angle for Vulcan XC72, Printex L6, 3% CeMnCo/VC, and 3% CeMnCo/PT, from left to right, respectively.

### 3.2. Oxygen Reduction Reaction (ORR)

Utilizing an RRDE, linear sweep voltammetry (LSV) was employed to conduct electrochemical analyses of ORR in an $O_2$-saturated alkaline electrolyte (1 mol $L^{-1}$ NaOH). The results for unmodified and modified carbon Vulcan XC72 in the absence and presence of a 2000-Oe magnetic field are exhibited in Fig. 7.

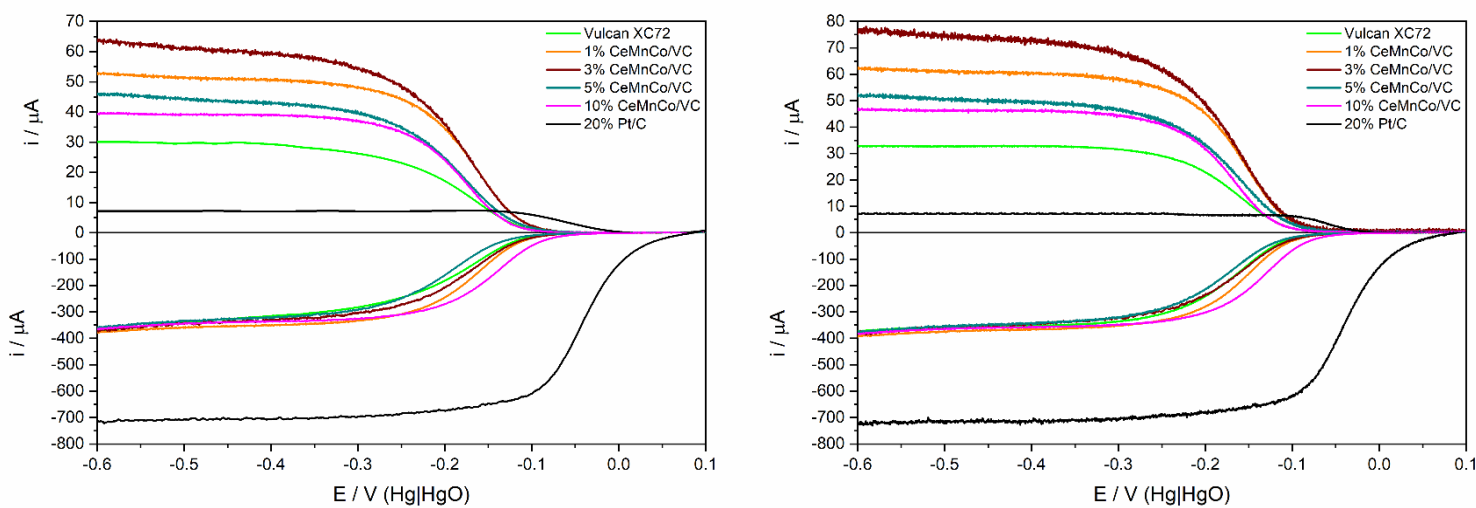

**Figure 7.** Steady-state polarization curves for CeMnCo/VC in the absence (left) and presence (right) of a 2000-Oe magnetic field (RRDE 1600 rpm and five mV s$^{-1}$).

Indeed, from Fig. 7, left, it is possible to realize that the modification of carbon Vulcan XC72 with CeMnCo nanoparticles enhanced the ring current (half the top of the figure). That is, the electrocatalyst's selectivity facilitates the 2-electron ORR, thus electrogenerating hydrogen peroxide [41].

Incorporating those nanoparticles in the Vulcan XC72 matrix led to a significant increase in the ring current, going from 31.0 (unmodified carbon) to 53.1, 64.0, 45.9, and 40.3 μA for 1, 3, 5, and 10% (w/w) CeMnCo. In other words, the ring current gains increase of 71.3, 106.4, 48.1, and 30.0 %. From these results, it is possible to observe an optimized quantity of the incorporation of nanoparticles in the carbonaceous matrix to reach a maximum point of efficiency, which in this case is 3% (w/w), suggesting a volcano-shape relationship [42].

With the search for innovative strategies in mind, utilizing magnetic fields in electrochemistry signifies a recent and inventive advancement. It demonstrates encouraging potential for enhancing diverse electrochemical reactions but is underexplored for the 2-electron ORR [43].

The experiments of RRDE were repeated, this time with a constant 2000-Oe magnetic field (Fig. 7-right). Results for unmodified Vulcan XC72 indicate a minimal effect on unmodified carbon, resulting in a relatively minor improvement in $H_2O_2$ selectivity. Given the marginal gain observed, it is probable that the magnetic field primarily affects the paramagnetic $O_2$ molecules rather than the carbonaceous matrix itself, inducing the Zeeman effect. The Zeeman effect is characterized by the interaction between the magnetic moment of an individual electron and an external magnetic field, leading to the splitting of energy levels associated with the unpaired electron [44, 45].

The incorporation of CeMnCo nanoparticles into the carbonaceous matrix significantly enhances the outcome, as illustrated by the results presented in Table 1.

**Table 1.** RRDE results from unmodified and modified carbon Vulcan XC72 in the absence and presence of a 2000-Oe magnetic field.

| Absence of a magnetic field | 2000-Oe magnetic field |
| --- | --- |

| Electrocatalyst | Disk current (µA) | Ring Current (µA) | Disk current (µA) | Ring Current (µA) |
|---|---|---|---|---|
| Vulcan XC72 | -375 | 31.0 | -373 | 32.7 |
| 1% CeMnCo/VC | -379 | 53.1 | -392 | 62.3 |
| 3% CeMnCo/VC | -373 | 64.0 | -375 | 75.3 |
| 5% CeMnCo/VC | -360 | 45.9 | -375 | 53.3 |
| 10% CeMnCo/VC | -365 | 40.3 | -381 | 46.6 |

When 1% (w/w) of CeMnCo is added to carbon Vulcan XC72, and a 2000-Oe magnetic field is used, the ring current is enhanced by around 17% compared to its absence. Also, the obtained result is almost twice that from unmodified carbon, evidencing the benefit of metal oxide addition to carbonaceous matrixes and using a magnetic field.

Utilizing the magnetic field for 3% CeMnCo/VC resulted in a substantial 18% augmentation of the ring current. Such enhancement is particularly significant compared with the unmodified Vulcan XC72, revealing a remarkable 2.3-fold increase in the achieved ring current. This observation underscores the synergistic impact of oxide nanoparticle incorporation and magnetic field manipulation, illuminating their collective potential to propel the efficiency of the two-electron oxygen reduction reaction.

The 5 and 10% CeMnCo/VC also presented increased results for ring current when the magnetic field was used (53.3 vs 45.9 µA and 46.6 vs 40.3 µA, respectively). When comparing against the unmodified carbonaceous matrix, the improvement went from 48.1 and 30.0% to 63.0 and 42.5%.

The determination of both $H_2O_2$ and $H_2O$ percentages and the assessment of the transferred electron number ($n_{e^-}$) were achieved using equations 1-3. Here, it represents the ring current, $i_d$ corresponds to the disk current, and N denotes the current collection efficiency of the Pt ring [46, 47]. The graphical representation of these with and without a magnetic field is depicted in Fig. 8.

$$H_2O_2\% = \frac{\frac{200 i_r}{N}}{i_d + \frac{i_r}{N}} \quad (1)$$

$$H_2O\% = 100 - H_2O_2\% \quad (2)$$

$$n_{e^-} = \frac{4i_d}{i_d + \frac{i_r}{N}} \qquad (3)$$

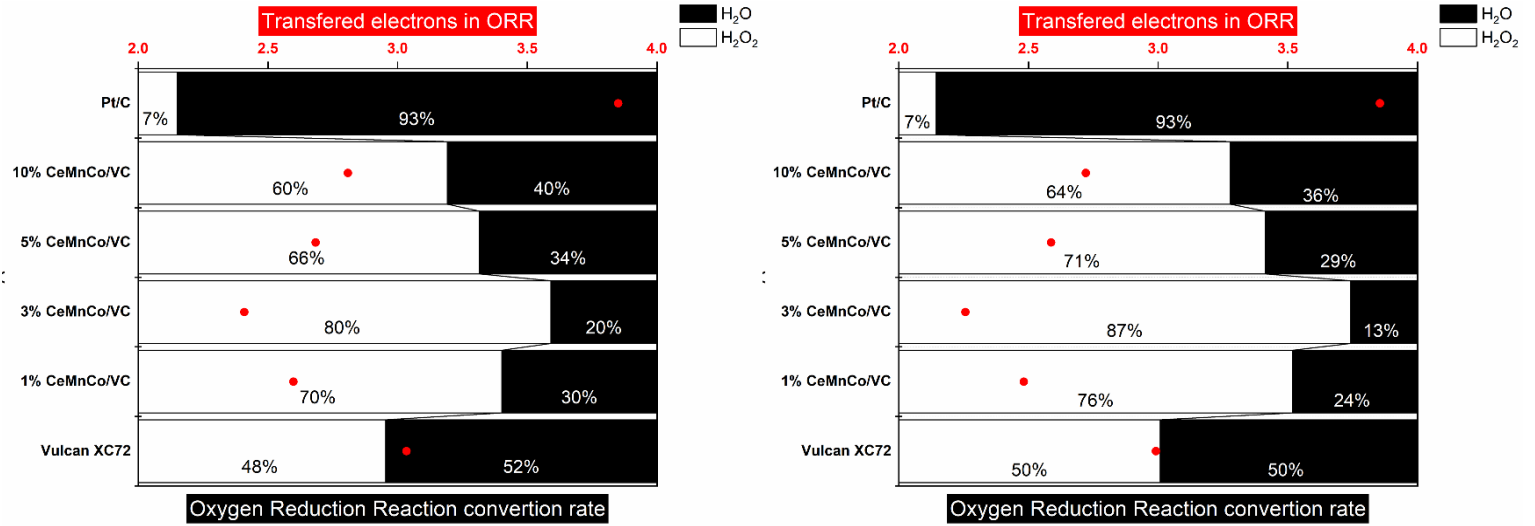

**Figure 8.** Electron transferred number and oxygen conversion rate on $H_2O_2/H_2O$ comparison for Vulcan XC72, CeMnCo/VC, and Pt/C in the absence (left) and presence (right) of a 2000-Oe magnetic field.

As expected, commercial platinum material favors the 4-electron oxygen reduction reaction, with $n_{e^-}$ being 3.9 and the selectivity of 93% for $H_2O$. The number of transferred electrons for the unmodified carbon Vulcan XC72 is 3.0 for both with and without the magnetic field. Despite that, there is a slightly better selectivity towards $H_2O_2$ in the experiments using the 2000-Oe magnetic field (50 vs 48%). This suggests that the Pauling adsorption model is somewhat favored because of the Zeeman effect the magnetic field causes on oxygen molecules.

The presence of CeMnCo in the carbon Vulcan XC72 matrix improved the 2-electron ORR. Without the magnetic field, the 10% (w/w) improved the $H_2O_2$ selectivity from 48% to 60%, with $n_{e^-}$ = 2.8. The 5% CeMnCo/VC was even better, with a selectivity of 66% towards hydrogen peroxide and a transferred electron number of 2.7. In the presence of 1% (w/w) of CeMnCo nanoparticles, the selectivity was boosted to 70% for $H_2O_2$ electrogeneration. The best result for $H_2O_2$ selectivity was achieved with 3% CeMnCo/VC, improving the 48% obtained with unmodified carbon Vulcan XC72 to 80%.

At 1% (w/w) CeMnCo loading, these nanoparticles play a crucial role as active sites. CeMnCo exhibits remarkable catalytic properties that are pivotal for facilitating the ORR, primarily through the 2-electron pathway. This unique attribute of CeMnCo significantly augments the catalytic activity of Vulcan XC72 with additional active sites and aids in oxygen adsorption and activation processes.

Furthermore, at 3% (w/w) CeMnCo nanoparticles, a substantial improvement in the 2e-ORR performance is observed. This enhancement stems from achieving a favorable equilibrium between active CeMnCo sites and the carbon support, thereby fostering synergistic interactions that facilitate efficient electron transfer and $O_2$ reduction, consequently enhancing overall catalytic activity.

However, excessive CeMnCo loading, such as at 5 and 10% w/w, may lead to diminished synergistic effects with carbon, resulting in nanoparticle agglomeration or clustering on the carbon surface. This phenomenon reduces active site accessibility and hinders efficient electron transfer, compromising the 2e-ORR activity. Nevertheless, these nanoparticles mixed into the carbon matrix are still beneficial for such a reaction when compared to unmodified Vulcan XC72 carbon.

Also, in Fig. 8, the 2000-Oe magnetic field substantially improves the $H_2O_2$ selectivity of all tested electrocatalysts. Results from modifying carbon Vulcan XC72 with CeMnCo nanoparticles were already great when combined with a magnetic field. The values were as follows:

- ∴ 1% CeMnCo/VC – from 70 to 76% (1.52-times unmodified Vulcan XC72 carbon).
- ∴ 3% CeMnCo/VC – from 80 to 87% (1.74-times unmodified Vulcan XC72 carbon).
- ∴ 5% CeMnCo/VC – from 66 to 71% (1.42 times unmodified Vulcan XC72 carbon).
- ∴ 10% CeMnCo/VC – from 60 to 64% (1.28 times unmodified Vulcan XC72 carbon).

These results show the importance of combining the modification of a carbonaceous matrix with metal oxide nanoparticles and a continuous magnetic field to improve $H_2O_2$ selectivity in the 2-electron ORR [48].

One consequence of a magnetic field is the Kelvin force ($F_K$). It arises due to the interaction between the magnetic field and polarized paired charge carriers free of unpaired charge carriers (electrons), inducing a force perpendicular to their velocity and the magnetic field direction. This force influences the movement and distribution of charged species in electrochemical reactions, potentially affecting the kinetics and selectivity of the ORR process. Moreover, the Kelvin force enhances the rapid movement of paramagnetic materials, boosting reaction rates near the electrode and guiding the flow of paramagnetic gases such as $O_2$ to areas with more vital magnetic fields [49].

The Lorentz force ($F_L$) induces macro- and microscopic convection by interacting with the local current density. Under the magnetic field's influence, magnetohydrodynamic effects induce a flow that thins the diffusion layer, enhancing the mass transport of reactants to the electrode surface. As a result, this causes the noted rise in current density during the oxygen reduction reaction. Additionally, the Lorentz force alters the trajectory and behavior of electrons participating in the ORR, potentially impacting its efficiency and selectivity towards $H_2O_2$ formation [50, 51].

The RRDE technique was also used to conduct electrochemical analyses of ORR in an $O_2$-saturated alkaline electrolyte (1 mol $L^{-1}$ NaOH) and evaluate the influence of CeMnCo nanoparticles on carbon Printex L6. The results are displayed in Fig. 9.

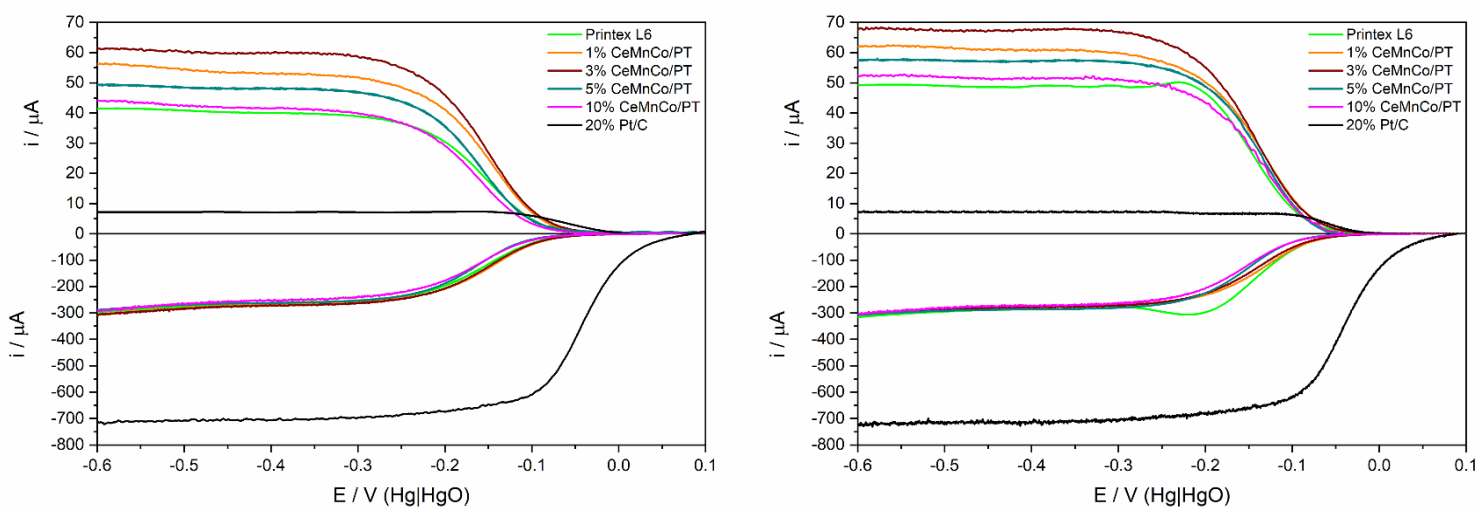

**Figure 9.** Steady-state polarization curves for CeMnCo/PT in the absence (left) and presence (right) of a 2000-Oe magnetic field (RRDE 1600 rpm and five mV s$^{-1}$).

Fig. 9 clearly shows that unmodified carbon Printex L6 is much better for $H_2O_2$ electrogeneration than unmodified carbon Vulcan XC72, with a ring current of 41.2 µA and a disk current of 295.0 µA versus a ring current of 31.0 µA and a disk current of 375.0 µA. These results corroborate those reported in the literature [52-54].

As the other carbonaceous matrix shows, incorporating CeMnCo nanoparticles for the carbon Printex L6 led to significant ring current enhancements. The values go from 41.1 to 56.8, 60.8, 49.3, and 44.1 µA for 1, 3, 5, and 10% (w/w) CeMnCo, representing gains of 38.2, 47.9, 20.0, and 8.0%. Once again, observations from our findings suggest a concentration-dependent relationship between the incorporation of nanoparticles into the carbonaceous matrix, highlighting a maximal efficiency point observed at 3% (w/w) concentration.

After doing the RRDE measurements with a 2000-Oe magnetic field (Fig. 9-suitable), a pronounced augment of the ring current in the unmodified carbon Printex L6 was observed, going from 41.1 to 49.2 µA, an improvement of 19.7%. The significant increase in ring current and consequent enhancement of $H_2O_2$ suggests a more pronounced effect of the magnetic field when compared to unmodified carbon Vulcan XC72. This enhanced response may be attributed to the structural or chemical characteristics of the Printex L6 carbonaceous matrix, which likely facilitate stronger interactions with paramagnetic $O_2$ molecules under the influence of the magnetic field in the Pauling model, leading to a more substantial improvement in $H_2O_2$ selectivity [55].

In summary, the differing responses of Vulcan XC72 and Printex L6 to the magnetic field can be attributed to their respective abilities to facilitate the adsorption of paramagnetic $O_2$ molecules under the Zeeman effect. This interaction ultimately impacts the efficiency of ORR and the selectivity towards $H_2O_2$ formation, highlighting the importance of the carbonaceous matrix in mediating the impact of external magnetic fields on catalytic processes.

Much like the effects on carbon Vulcan XC72, by integrating CeMnCo nanoparticles into the carbon Printex L6 matrix, the impact of a magnetic field becomes notably more prominent, as evidenced by the outcomes delineated in Table 2.

**Table 2.** RRDE results from unmodified and modified carbon Printex L6 without and with a 2000-Oe magnetic field.

| Electrocatalyst | Absence of a magnetic field | | 2000-Oe magnetic field | |
|---|---|---|---|---|
| | Disk current (µA) | Ring Current (µA) | Disk current (µA) | Ring Current (µA) |
| **Printex L6** | -295 | 41.2 | -316 | 49.2 |
| **1% CeMnCo/PT** | -308 | 56.8 | -303 | 61.8 |
| **3% CeMnCo/PT** | -307 | 60.8 | -305 | 67.5 |
| **5% CeMnCo/PT** | -288 | 49.3 | -309 | 57.6 |
| **10% CeMnCo/PT** | -294 | 44.1 | -302 | 52.3 |

Introducing a magnetic field resulted in an 8.8% enhancement in the ring current of the 1% CeMnCo/PT sample compared to its absence. It corresponded to a 1.25-fold increase relative to unmodified carbon Printex L6. The 3% CeMnCo/PT sample remained the most effective electrocatalyst. Furthermore, applying a 2000-Oe magnetic field led to an approximate 11% enhancement in ring current compared to its absence, translating to a 1.27-fold increase compared to unmodified carbon Printex L6.

In the RRDE measurements, incorporating a magnetic field notably boosted the ring current for the 5% and 10% CeMnCo/PT electrocatalysts, marking increments of 16.8% and 18.6%, respectively. Comparatively, about unmodified carbon Printex L6, the presence of nanoparticles within the carbonaceous matrix coupled with 2000-Oe magnetic field yielded enhancements of 1.17-fold and 1.06-fold for the 5% and 10% CeMnCo/PT electrocatalysts, respectively.

Again, Fig. 10 illustrates the percentages of hydrogen peroxide and water and evaluates the transferred electron number ($n_{e^-}$) parameters under both conditions, with and without the presence of a magnetic field.

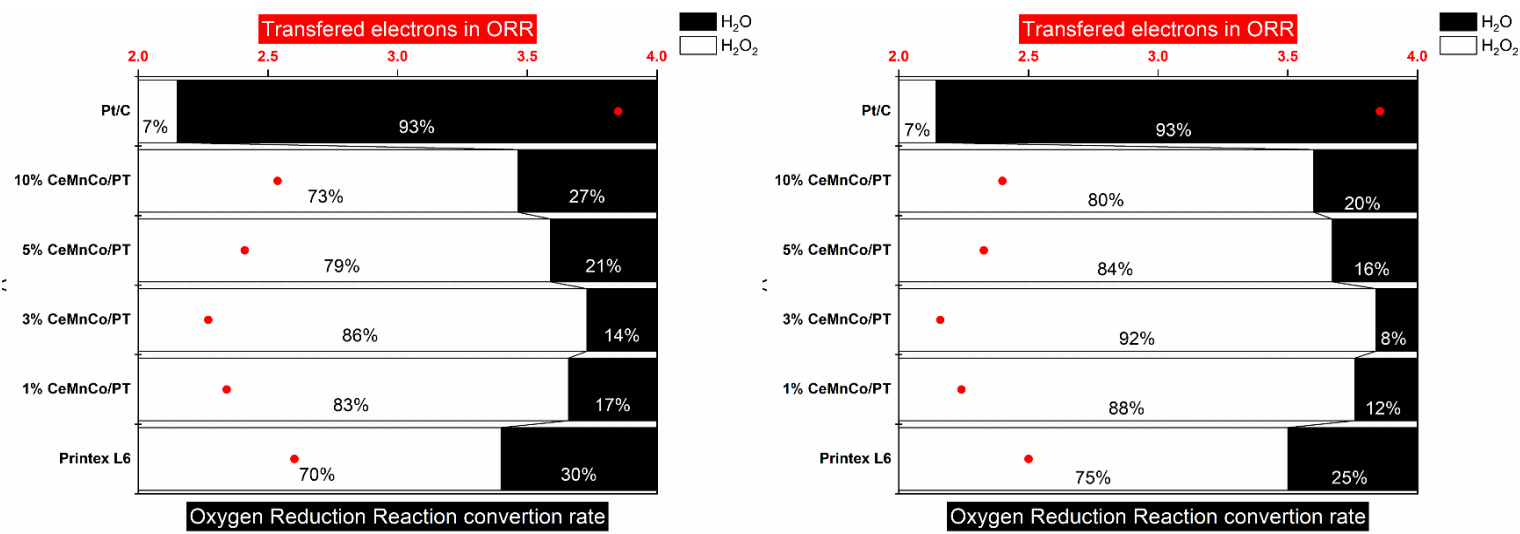

**Figure 10.** Electron transferred number and O$_2$ conversion rate on H$_2$O$_2$/H$_2$O comparison for Printex L6, CeMnCo/PT, and Pt/C without (left) and with (right) a 2000-Oe magnetic field.

In the case of unmodified carbon Printex L6, applying a 2000-Oe magnetic field induces a shift in the number of transferred electrons from 2.6 to 2.5. This phenomenon is attributed to the Zeeman effect observed in O$_2$ paramagnetic molecules, which alters electron transfer dynamics [56]. Consequently, the magnetic field promotes the Pauling adsorption model on the surface of the carbon electrocatalyst, thereby enhancing the selectivity for H$_2$O$_2$ electrogeneration during the 2-electron ORR. This enhancement is evidenced by an increase in selectivity from 70% to 75%.

Unmodified carbon Printex L6 catalyzes the 2-electron ORR and H$_2$O$_2$ electrogeneration. So, modifying its matrix and improving the hydrogen peroxide selectivity is difficult. One that was achieved with CeMnCo nanoparticles, as shown in Fig. 10.

With CeMnCo 10% (w/w), the H$_2$O$_2$ selectivity enhanced to 73% and 80%, without and with the magnetic field, improving $n_{e^-}$ to 2.5 and 2.4. Even better results were achieved with the CeMnCo 5% (w/w) electrocatalyst. The selectivity without the magnetic field increased from 70 to 79%, accompanied by a change in the number of transferred electrons from 2.6 to 2.4. And with a 2000-Oe magnetic field, the selectivity for H$_2$O$_2$ electrogeneration rose from 75 to 84%, while $n_{e^-}$ decreased from 2.5 to 2.3.

With CeMnCo 1% (w/w) modification, the $H_2O_2$ selectivity increased to 83 and 88%, respectively, in the absence and presence of a 2000-Oe magnetic field, with the number of transferred electrons decreasing from 2.3 to 2.2. Remarkably, the most significant improvements were observed with CeMnCo 3% (w/w) modifying the Printex L6 matrix. This modification led to a remarkable enhancement in hydrogen peroxide electrogeneration from 86 to 92%, respectively, without and with a 2000-Oe magnetic field, coinciding with a decrease in the number of transferred electrons from 2.3 to 2.2.

Once again, our findings underscore the significance of integrating the modification of a carbonaceous matrix with metal oxide nanoparticles alongside a consistent magnetic field, which demonstrates enhanced $H_2O_2$ selectivity during the 2-electron ORR process. The theory and explanation regarding ceria nanostructures published by our research group [48] can be extended to $Ce_{1.0}Mn_{0.9}Co_{0.1}$ nanoparticles modified on a carbonaceous matrix for electrocatalysis in the presence of a 2000-Oe magnetic field to enhance ring current in RRDE and improve $H_2O_2$ selectivity.

In this context, oxygen vacancies within the $Ce_{1.0}Mn_{0.9}Co_{0.1}$ nanoparticles, confirmed by magnetization curves, could act as an oxygen buffer, similar to ceria nanostructures reported in the literature [57]. Cerium oxide, including the modified nanoparticles, exhibits susceptibility to magnetic fields due to ferromagnetic ordering arising from unpaired electrons of $Ce^{3+}$ ions and oxygen vacancies. When subjected to a magnetic field, an energy level split occurs due to spin interactions between $Ce^{3+}$ ions and oxygen vacancies, forming polarons and bound magnetic polarons (BMPs). BMPs are clusters of spins organized ferromagnetic within the Bohr radius, facilitated by exchange interactions with effective mass carriers in localized states [48].

Moreover, the presence of manganese in the $Ce_{1.0}Mn_{0.9}Co_{0.1}$ nanoparticles introduces additional magnetic properties. Manganese ions can exhibit diverse magnetic behaviors depending on their oxidation states and local coordination environment. In this case, the presence of manganese can contribute to the overall magnetic response of the nanoparticles. Under the influence of the applied magnetic field, manganese ions may also experience energy level shifts and spin interactions, like cerium ions and oxygen vacancies. These magnetic effects of manganese, combined with those of cerium and oxygen vacancies, further contribute to forming BMPs and enhancing electron spin states within the nanoparticles. Additionally, the presence of Co and Mn in $Ce_{1.0}Mn_{0.9}Co_{0.1}$ nanoparticles enhances the generation of oxygen vacancies by substituting Ce ions with

Mn ions on the nanoparticle surface [58] or by initiating vacancy nucleation in the vicinity of Co ions.

Therefore, the application of a magnetic field enhances the energy states of electron spins within the $Ce_{1.0}Mn_{0.9}Co_{0.1}$ nanoparticles, like ceria nanostructures. This enhancement can facilitate the catalytic reaction, including the 2e-ORR, by promoting efficient electron transfer and potentially improving selectivity towards $H_2O_2$ formation. Additionally, the presence of oxygen vacancies, cerium, and manganese ions, along with the formation of BMPs, collectively contribute to the observed enhancement in catalytic activity and selectivity, particularly in the context of ring current enhancement in RRDE experiments.

## 4. Conclusions

The present study demonstrated the successful synthesis of Ce1.0Mn0.9Co0.1 nanoparticles with XRD, EPR, magnetization curves, and TEM/HRTEM/EDX. This material was mixed with carbon Vulcan XC72 and carbon Printex L6 in different proportions (1, 3, 5, and 10 % w/w) to electrocatalysts for the 2-electron oxygen reduction reaction. Before the tests, the material was also characterized via XPS and contact angle. Our electrochemical studies produced convincing findings, demonstrating that the nanostructures altered both carbonaceous matrices across various concentrations (1, 3, 5, and 10% w/w) towards the 2-electron ORR. This modification significantly boosted ring currents in RRDE measurements, indicating improved selectivity for $H_2O_2$ production. Our research demonstrated the significant impact of Magnetic Field-Enhanced Electrochemistry, utilizing a continuous magnetic field strength of 2000 Oe, on 2-electron ORR experiments. Notably, the results exceeded those obtained without the magnetic field, highlighting increased ring currents and enhanced selectivity for $H_2O_2$ production when using CeMnCo nanoparticles. These notable improvements in electrocatalytic performance were linked to the influence of the magnetic field, which was connected to the Zeeman effect in $O_2$ molecules and the interaction of Lorentz and Kelvin forces in conjunction with the Bound Magnetic Polarons within the nanostructures and the enhanced generation of oxygen vacancies.

**Credit authorship contribution statement**

**Caio Machado Fernandes:** Investigation, Validation, Data curation, Writing – original draft. **João Paulo C. Moura:** Validation, Writing – review & editing. **Aline B. Trench:** Data curation, Writing – original draft. **Odivaldo C. Alves:** Writing – original draft and review & editing. **Yutao Xing:** Investigation, Validation, Writing – original draft and review & editing, **Marcos R. V. Lanza:** Writing – review & editing. **Júlio César M. Silva:** Writing – review & editing. **Mauro C. Santos:** Conceptualization, Writing – review & editing, Supervision.

**Data availability**

The raw/processed data required to reproduce these findings cannot be shared due to legal or ethical reasons.

**Declaration of Competing Interest**

The authors declare that they have no known competing financial interests or personal relationships that could have appeared to influence the work reported in this paper.

**Acknowledgments**

The authors would like to thank Fundação de Amparo à Pesquisa do Estado de São Paulo (FAPESP, #2021/05364-7, #2021/14394-7, #2022/10484-4, #2022/12895-1, and #2022/15252-4) for the financial support. The authors are also grateful for Coordenação de Aperfeiçoamento de Pessoal de Nível Superior (CAPES) and Conselho Nacional de Desenvolvimento Científico e Tecnológico (CNPq) (#303943/2021-1, #308663/2023–3, #402609/2023–9) for their support.